\documentclass[smallextended]{svjour3}       
\smartqed  
\usepackage{graphicx,epsfig}

\begin{document}

\title{
Local quantum Fisher information and local quantum uncertainty
for general X-states
}

\author{
M.~A.~Yurischev\and~Saeed~Haddadi
}


\institute{
M.~A.~Yurischev
\at
Federal Research Center of Problems of Chemical Physics and Medicinal Chemistry,
Russian Academy of Sciences,
Chernogolovka 142432, Moscow Region, Russia\\     
\email{yur@itp.ac.ru}
           \and
S. Haddadi \at 
School of Physics, Institute for Research in Fundamental Sciences (IPM), P.O. Box
19395-5531, Tehran, Iran\\
and\\
Saeed's Quantum Information Group, P.O. Box 19395-0560, Tehran, Iran\\
              \email{saeed@ssqig.com}           
}

\date{Received:}

\titlerunning{
Local quantum Fisher information and local quantum uncertainty
}
\maketitle

\begin{abstract}
A two-spin-1/2 Heisenberg XYZ model with Dzyaloshinsky--Moriya (DM) and
Kaplan--Shekhtman--Entin-Wohlman--Aharony (KSEA) interactions in the presence of an
inhomogeneous external magnetic field is considered at thermal equilibrium.
Its density matrix has the general X form for which we derive explicit formulas for
the local quantum Fisher information (LQFI) and local quantum uncertainty (LQU)
directly in terms of model parameters.
This allows us to perform a comparative study for the discord-type quantum
correlations LQFI and LQU and to reveal a number of new features in their behavior.
In particular, the sudden transitions of quantum correlations with a smooth change in
temperature are found.
Moreover, it is shown that there may be a sequence of such transitions for certain
choices of interaction constants.
\end{abstract}


\section{Introduction}
\label{sect:Intro}
For a long time, the phenomenon of entanglement was considered to be the only
ingredient of quantum properties.
Quantum entangled states play a critical role in quantum cryptography, quantum
computing, superdense coding, quantum teleportation, etc.
\cite{GRTZ02,V05,V05a,AFOV08,HHHH09}.

In quantum informatics, entanglement is viewed as a physical resource
(``as real as energy'' \cite{HHHH09})\footnote{
 However, ``it is important to realize that in physics today, we have no knowledge of
 what the energy {\em is}'' \cite {FLS64}, Sect.~4-1.
}.
This statement requires a mathematical definition, since, according to Kant,
``in jeder besonderen Naturlehre nur so viel eigentliche Wissenschaft angetroffen
werden k$\rm\ddot o$nne, als darin Mathematik anzutreffen ist''
(``in any special doctrine of nature there can be only as much {\em proper\/} science
as there is {\em mathematics} therein'').
Quantum entanglement was quantified in 1996, first for pure states
\cite{BBPSSW96,BBPS96}, and then for mixed states \cite{BDSW96}.
According to the accepted definition, the entanglement of a bipartite pure state is
the von Neumann entropy either of the two subsystems.\footnote{
 Earlier, a similar definition was proposed by Everett for the
 {\em canonical correlation} \cite{E73}.
}
The entanglement (of formation) of a bipartite mixed state is defined as the
minimum entanglement of an ensemble over all ensembles realizing the mixed state.
This gave impetus to quantifying entanglement (and then other quantum correlations)
based on various criteria \cite{GMPZ22}.

Despite the fundamental and practical importance of quantum entanglement, subsequent
studies have shown that it does not exhaust all quantum correlations in a system.
Using the model of deterministic quantum computation with one pure qubit (DQC1), Knill
and Laflamme showed that computation can achieve an exponential improvement in
efficiency over classical computers even without containing much entanglement
\cite{KL98}.

In 2000, $\rm\dot Z$urek developed the concept of quantum discord \cite{Z00}.
Within this concept, discord was interpreted as
``a measure of the quantumness of correlations'' \cite{OZ01}.
The quantum discord ${\cal Q}$ for a bipartite system $AB$ is defined as the minimum
difference between the quantum generalizations of symmetric ($I$) and asymmetric
looking ($J$) versions of the classical mutual information:
${\cal Q}=\min_{\{{\rm\Pi}_k\}}(I-J)$, where $\{{\rm\Pi}_k\}$ is the measurement
performed on {\em one} of the two subsystems \cite{Z00,OZ01}.

Quantum discord equals zero for the classically correlated states and coincides the
entropy of entanglement for the pure states.
It is remarkable that discord can exist even in separable (but mixed) states, i.e.,
when quantum entanglement is identically equal to zero.
Thus, quantum discord is a different measure of quantum correlation than entanglement.
Datta et al. \cite{D08,DSC08} calculated discord in the Knill-Laflamme DQC1 model
and showed that it scales with the quantum efficiency, while entanglement remains
vanishingly small throughout the computation.
This attracted a lot of attention to the new measure of quantum correlation
\cite{MBCPV12,AFY14,AFY14a,S15}.

The notion of quantum discord is based on local measurements and optimization.
This idea was then applied to other physical or information quantities,
and to date, a large number of discord-like measures of quantum correlation have been
introduced \cite{MBCPV12,ABC16,FSA17,BDSRSS18}.
The local quantum uncertainty (LQU) \cite{GTA13} and local quantum Fisher information
(LQFI) \cite{GSGTFSSOA14,B14,KLKW18} belong to this type of measures.

The behavior of LQFI and LQU in various specific two-qubit systems was considered by
many researchers
\cite{SBDL19,H20,YLLF20,MC21,BSHD22,KDS19,BAD22,EDM22,BRHD22,BHAPDA22,DRM23}.
In the previous paper \cite{FY22}, the behavior of LQU and LQFI (together with the
entropic discord) was studied  on the example of Bell-diagonal states, which are a
subclass of X states and allow to include both DM and KSEA interactions but without
Zeeman's terms.
Surprisingly, an excellent agreement of these quantifiers of quantum correlation
was observed.
In the present paper, we extend this consideration to general X states and
describe new features and peculiarities in the behavior of LQFI and LQU.

The structure of the paper is as follows.
Section~\ref{sect:H-rho} presents the Hamiltonian and the corresponding Gibbs density
matrix; the diagonalization of the density matrix is also described here.
Derivation of formulas for the LQFI and LQU is given in Sects.~\ref{sect:LQFI} and
\ref{sect:LQU}, respectively. 
In Sect.~\ref{sect:Discus} we present the results and discuss them.
Finally, Sect.~\ref{sect:Concl} summarizes our findings.

\section{
Hamiltonian, density matrix and its diagonalization
}
\label{sect:H-rho}
Let there be a system with some Hamiltonian.
Then its density operator (matrix)
is found, say, from the quantum Liouville-von Neumann or Lindblad master equation, or
has a thermal equilibrium Gibbs’ form.
Here we restrict ourselves to the latter case.

Consider a model with the following Hamiltonian \cite{Y20}
\begin{eqnarray}
   \label{eq:H}
   H&=&
	 J_x\sigma^1_x\sigma^2_x + J_y\sigma^1_y\sigma^2_y + J_z\sigma^1_z\sigma^2_z
	 +D_z(\sigma^1_x\sigma^2_y-\sigma^1_y\sigma^2_x)
	 +{\rm \Gamma}_z(\sigma^1_x\sigma^2_y+\sigma^1_y\sigma^2_x)
	 \nonumber\\
   &+&B_1\sigma^1_z + B_2\sigma^2_z,
\end{eqnarray}
where $\sigma^i_\alpha$ ($i=1,2$; $\alpha=x,y,z$) are the Pauli spin operators,
($J_x$,$J_y$,$J_z$) the vector of interaction constants of the Heisenberg part of
interactions,
$D_z$ the $z$-component of Dzyaloshinsky vector,
${\rm\Gamma}_z$ the strength of KSEA interaction, and
$B_1$ and $B_2$ the $z$-components of external magnetic fields applied at the first
and second spins, respectively.
Thus, this model contains seven real independent parameters: $J_x$,
$J_y$, $J_z$, $D_z$, ${\rm\Gamma}_z$ $B_1$, and $B_2$.

In open form, the Hamiltonian (\ref{eq:H}) reads
\begin{equation}
   \label{eq:Hm}
   H=
	 \left(
      \begin{array}{cccc}
      J_z+B_1+B_2&.&.&J_x-J_y-2i{\rm \Gamma}_z\\
      .&-J_z+B_1-B_2\ &J_x+J_y+2iD_z&.\\
      .&J_x+J_y-2iD_z\ &-J_z-B_1+B_2&.\\
      J_x-J_y+2i{\rm \Gamma}_z&.&.&J_z-B_1-B_2
      \end{array}
   \right)\!\!,
\end{equation}
with the dots which were put instead of zero entries.
This is the most general Hermitian traceless matrix having the X structure: $(\times)$;
that is its nonzero entries may belong only to the main diagonal and anti-diagonal.
Note that the set of X matrices is algebraically closed: their sums and products are
again the X matrices.

Eigenlevels of the Hamiltonian under consideration are given by
\begin{equation}
   \label{eq:Ei}
   E_{1,2}=J_z\pm R_1,\qquad E_{3,4}=-J_z\pm R_2,
\end{equation}
where
\begin{equation}
   \label{eq:R1R2}
   R_1=[r_1^2+(B_1+B_2)^2]^{1/2},\quad
   R_2=[r_2^2+(B_1-B_2)^2]^{1/2},
\end{equation}
with
\begin{equation}
   \label{eq:R1R2b}
   r_1=[(J_x-J_y)^2+4{\rm\Gamma}_z^2]^{1/2},\quad
   r_2=[(J_x+J_y)^2+4D_z^2]^{1/2}.
\end{equation}

The Gibbs density matrix is written as
\begin{equation}
   \label{eq:rho}
   \rho=\frac{1}{Z}\exp(-\beta H).
\end{equation}
Here, the partition function $Z=\sum_n\exp(-\beta E_n)$ is expressed as
\begin{equation}
   \label{eq:Z}
   Z=2[e^{-\beta J_z}\cosh(\beta R_1)+e^{\beta J_z}\cosh(\beta R_2)],
\end{equation}
where $\beta=1/T$, with $T$ being the temperature in energy units.
The operator $\rho$ satisfies the conditions:
$\rho^\dagger=\rho$, $\rho\ge0$, and ${\rm tr}\rho=1$.

Due to the functional relation (\ref{eq:rho}) and algebraic
closeness of X matrix set, the Gibbs density matrix has the following X form:
\begin{equation}
   \label{eq:rho-zz}
   \rho=
	 \left(
      \begin{array}{cccc}
      a&.&.&u\\
      .&b\ &v&.\\
      .&v^*\ &c&.\\
      u^*&.&.&d
      \end{array}
   \right)\!\!,
\end{equation}
where the asterisk denotes complex conjugation.
Performing the necessary calculations, we obtain expressions for the matrix elements of
$\rho$:
\begin{eqnarray}
   \label{eq:avT}
   &&a=\frac{1}{Z}\{\cosh(\beta R_1)-[(B_1+B_2)/R_1]\sinh(\beta R_1)\}e^{-\beta J_z},
   \nonumber\\
   &&b=\frac{1}{Z}\{\cosh(\beta R_2)-[(B_1-B_2)/R_2]\sinh(\beta R_2)\}e^{\beta J_z},
   \nonumber\\
   &&c=\frac{1}{Z}\{\cosh(\beta R_2)+[(B_1-B_2)/R_2]\sinh(\beta R_2)\}e^{\beta J_z},\\
   &&d=\frac{1}{Z}\{\cosh(\beta R_1)+[(B_1+B_2)/R_1]\sinh(\beta R_1)\}e^{-\beta J_z},
   \nonumber\\
   &&u=-\frac{1}{Z}[(J_x-J_y-2i{\rm\Gamma}_z)/R_1]\sinh(\beta R_1)e^{-\beta J_z},
   \nonumber\\
   &&v=-\frac{1}{Z}[(J_x+J_y+2iD_z)/R_2]\sinh(\beta R_2)e^{\beta J_z},
   \nonumber
\end{eqnarray}
where $R_1$ and $R_2$ are given again by Eq.~(\ref{eq:R1R2}).

Using the invariance of quantum correlations under any local unitary transformations,
we remove complex phases in the off-diagonal entries and change $\rho\to\varrho$, where
\begin{eqnarray}
   \label{eq:varrho}
   \varrho=
	 \left(
      \begin{array}{cccc}
      a&.&.&|u|\\
      .&b&|v|&.\\
      .&|v|&c&.\\
      |u|&.&.&d
      \end{array}
   \right)\!\!,
\end{eqnarray}
with
\begin{equation}
   \label{eq:avTm}
   |u|=\frac{r_1\sinh(\beta R_1)}{ZR_1}e^{-\beta J_z},\qquad
   |v|=\frac{r_2\sinh(\beta R_2)}{ZR_2}e^{\beta J_z}.
\end{equation}
Due to the conditions of Hermitianity,
non-negativity, and normalization
of any density operator, the matrix elements $a,b,c,d\ge0$, $a+b+c+d=1$, $ad\ge|u|^2$,
and $bc\ge|v|^2$.
The quantum state $\varrho$ depends on five real parameters, for which
we can take $J_z$, $R_1$, $R_2$, $B_1$, and $B_2$ or, say, $J_z$, $r_1$, $r_2$, $B_1$,
and $B_2$.
Note that the quantum state (\ref{eq:varrho}) is reduced to the Bell-diagonal case
when $a=d$ and $b=c$, i.e., when $B_1=B_2=0$.

Transform now the density matrix $\varrho$ into diagonal representation.
We will perform this transformation in two steps.
First, using symmetry of any X matrix with respect to the transformations of the group
$\{E,\sigma_z\otimes\sigma_z\}$ \cite{Y20}, one can reduce
the matrix (\ref{eq:varrho}) to the block-diagonal form.
It can be achieved by a simultaneous permutation of 2-nd and 4-th rows and columns of
X matrix:
\begin{eqnarray}
   \label{eq:varrho1}
   P\varrho P^t=
	 \left(
      \begin{array}{cccc}
      a&|u|&.&.\\
      |u|&d&.&.\\
      .&.&c&|v|\\
      .&.&|v|&b
      \end{array}
   \right),
\end{eqnarray}
where
\begin{eqnarray}
   \label{eq:P}
   P=
	 \left(
      \begin{array}{cccc}
      1&.&.&.\\
      .&.&.&1\\
      .&.&1&.\\
      .&1&.&.
      \end{array}
   \right)=P^t.
\end{eqnarray}
Here the subscript $t$ stands for matrix transpose.
From Eq.~(\ref{eq:varrho1}), it is clear that the eigenvalues of $\varrho$ are equal to
\begin{equation}
   \label{eq:p_i}
   p_{1,2}=\frac{1}{2}\Big(a+d\pm\sqrt{(a-d)^2+4|u|^2}\Big),\quad
	 p_{3,4}=\frac{1}{2}\Big(b+c\pm\sqrt{(b-c)^2+4|v|^2}\Big).
\end{equation}

Second, we now use orthogonal transformation (it is built from eigenvectors
of the subblocks $2\times2$; see, e.g., \cite{M57})
\begin{equation}
   \label{eq:R}
   R=
	 \left(
      \begin{array}{ccrr}
      q_1/\sqrt{q_1^2+|u|^2}&|u|/\sqrt{q_1^2+|u|^2}&.&.\\
      |u|/\sqrt{q_1^2+|u|^2}&-q_1/\sqrt{q_1^2+|u|^2}&.&.\\
      .&.&q_2/\sqrt{q_2^2+|v|^2}&|v|/\sqrt{q_2^2+|v|^2}\\
      .&.&|v|/\sqrt{q_2^2+|v|^2}&-q_2/\sqrt{q_2^2+|v|^2}
      \end{array}
   \right)=R^t,
\end{equation}
where
\begin{equation}
   \label{eq:q}
   q_1=\frac{1}{2}\Big(a-d+\sqrt{(a-d)^2+4|u|^2}\Big),\quad
   q_2=\frac{1}{2}\Big(c-b+\sqrt{(c-b)^2+4|v|^2}\Big).
\end{equation}
Note the useful expressions:
\begin{equation}
   \label{eq:q2}
   q_1^2=(a-d)q_1+|u|^2,\quad
   q_2^2=(c-b)q_2+|v|^2.
\end{equation}
Here, one should keep the carefulness when $|u|$ or $|v|$ equals zero.

As a result,
\begin{eqnarray}
   \label{eq:RPvarrhoPR}
   RP\varrho PR=
	 \left(
      \begin{array}{cccc}
      p_2&.&.&.\\
      .&p_1\ &.&.\\
      .&.\ &p_4&.\\
      .&.&.&p_3
      \end{array}
   \right),
\end{eqnarray}
where $p_i$ are still the eigenvalues (with an insignificant permutation) of the matrix
(\ref{eq:varrho}).

Now, using transformations (\ref{eq:P}) and (\ref{eq:R}), we find local spin matrices
$\sigma_\mu\otimes I$ ($\mu=x,y,z$) in the diagonal representation of the density
matrix $\varrho$ (i.e, we get the sets of matrix elements
$\langle m|\sigma_\mu\otimes I|n\rangle$):
\begin{eqnarray}
   \label{eq:RPxPR}
   &&RP(\sigma_x\otimes I)PR=\frac{1}{\sqrt{(q_1^2+|u|^2)(q_2^2+|v|^2)}}
	 \nonumber\\
	 &&\times
	 \left(
      \begin{array}{cccc}
      .&.&q_1q_2+|uv|&q_1|v|-q_2|u|\\
      .&.&q_2|u|-q_1|v|&q_1q_2+|uv|\\
      q_1q_2+|uv|&q_2|u|-q_1|v|&.&.\\
      q_1|v|-q_2|u|&q_1q_2+|uv|&.&.
      \end{array}
   \right)\!\!,\qquad 
\end{eqnarray}
\begin{eqnarray}
   \label{eq:RPyPR}
   &&RP(\sigma_y\otimes I)PR=\frac{i}{\sqrt{(q_1^2+|u|^2)(q_2^2+|v|^2)}}
	 \nonumber\\
	 &&\times
	 \left(
      \begin{array}{cccc}
      .&.&|uv|-q_1q_2&-q_1|v|-q_2|u|\\
      .&.&-q_1|v|-q_2|u|&q_1q_2-|uv|\\
      q_1q_2-|uv|&q_1|v|+q_2|u|&.&.\\
      q_1|v|+q_2|u|&|uv|-q_1q_2&.&.
      \end{array}
   \right)\!\!\qquad 
\end{eqnarray}
and
\begin{eqnarray}
   \label{eq:RPzPR}
   RP(\sigma_z\otimes I)PR=
	 \left(
      \begin{array}{cccc}
      \frac{q_1(a-d)}{q_1^2+|u|^2}&\frac{2q_1|u|}{q_1^2+|u|^2}&.&.\vspace{1mm}\\ 
      \frac{2q_1|u|}{q_1^2+|u|^2}&\frac{q_1(d-a)}{q_1^2+|u|^2}&.&.\\
      .&.&\frac{q_2(b-c)}{q_2^2+|v|^2}&\frac{-2q_2|v|}{q_2^2+|v|^2}\vspace{1mm}\\ 
      .&.&\frac{-2q_2|v|}{q_2^2+|v|^2}&\frac{q_2(c-b)}{q_2^2+|v|^2}
      \end{array}
   \right)\!\!.\ 
\end{eqnarray}
So, now everything is ready to start calculating LQFI and LQU.

\section{
Local quantum Fisher information
}
\label{sect:LQFI}
The LQFI measure is based on the quantum Fisher information $F$ \cite{H82,BC94}.
It was suggested in Ref.~\cite{GSGTFSSOA14} (see there especially Supplementary
Information); see also \cite{B14,KLKW18}.
This measure which we will denote by $\cal F$ equals the optimal LQFI with the
measuring operator $H_A$ acting in the subspace of party $A$ of the bipartite system
$AB$:
\begin{equation}
   \label{eq:Fdef}
   {\cal F}(\varrho)=\min_{H_A}F(\varrho,H_A).
\end{equation}

Remarkably, the authors \cite{GSGTFSSOA14} proved that, if the subsystem $A$ is a
qubit, the optimization in Eq.~(\ref{eq:Fdef}) can be performed and 
LQFI is
\begin{equation}
   \label{eq:Fm}
   {\cal F}=1-\lambda_{max}^{(M)},
\end{equation}
where $\lambda_{max}^{(M)}$ is the largest eigenvalue of the real symmetric $3\times3$
matrix $M$ with entries (see also \cite{DBA15,MKYE21,MKHRTP22,ZRHCH23})
\begin{equation}
   \label{eq:M}
   M_{\mu\nu}=\sum_{\scriptstyle m,n\atop\scriptstyle p_m+p_n\ne0}\frac{2p_mp_n}{p_m+p_n}\langle m|\sigma_\mu\otimes I|n\rangle
	 \langle n|\sigma_\nu\otimes I|m\rangle.
\end{equation}
Using Eqs.~(\ref{eq:RPxPR})--(\ref{eq:RPzPR}), we find that the matrix $M$ is
diagonal and its eigenvalues are equal to
\begin{eqnarray}
   \label{eq:Mxx0}
   M_{xx}&=&\frac{4}{(q_1^2+|u|^2)(q_2^2+|v|^2)}\Biggl[(q_1q_2+|uv|)^2\Bigg(\frac{p_1p_3}{p_1+p_3}+\frac{p_2p_4}{p_2+p_4}\Bigg)
	 \nonumber\\
	 &+&(q_1|v|-q_2|u|)^2\Biggl(\frac{p_1p_4}{p_1+p_4}+\frac{p_2p_3}{p_2+p_3}\Biggr)\Bigg],
\end{eqnarray}
\begin{eqnarray}
   \label{eq:Myy0}
   M_{yy}&=&\frac{4}{(q_1^2+|u|^2)(q_2^2+|v|^2)}\Biggl[(q_1q_2-|uv|)^2\Bigg(\frac{p_1p_3}{p_1+p_3}+\frac{p_2p_4}{p_2+p_4}\Bigg)
	 \nonumber\\
	 &+&(q_1|v|+q_2|u|)^2\Biggl(\frac{p_1p_4}{p_1+p_4}+\frac{p_2p_3}{p_2+p_3}\Biggr)\Bigg]
\end{eqnarray}
and
\begin{eqnarray}
   \label{eq:Mzz0}
   M_{zz}&=&\frac{1}{(q_1^2+|u|^2)^2}\Bigg[(a+d)(q_1^2-|u|^2)^2+16\frac{ad-|u|^2}{a+d}q_1^2|u|^2\Bigg]
	 \nonumber\\
   &+&\frac{1}{(q_2^2+|v|^2)^2}\Bigg[(b+c)(q_2^2-|v|^2)^2+16\frac{bc-|v|^2}{b+c}q_2^2|v|^2\Bigg],
\end{eqnarray}
where $q_1$ and $q_2$ are given by Eq.~(\ref{eq:q}) and $p_1$ to $p_4$ are determined by
Eq.~(\ref{eq:p_i}).

Through tedious calculations, we arrive at expressions for the eigenvalues of matrix
$M$:
\begin{equation}
   \label{eq:Mxx}
   M_{xx}=\frac{64(ac+bd+p_1p_2+p_3p_4+2|uv|)[(a+d)p_3p_4+(b+c)p_1p_2]}{[1-(p_1-p_2)^2-(p_3-p_4)^2]^2-4(p_1-p_2)^2(p_3-p_4)^2},
\end{equation}
\begin{equation}
   \label{eq:Myy}
   M_{yy}=\frac{64(ac+bd+p_1p_2+p_3p_4-2|uv|)[(a+d)p_3p_4+(b+c)p_1p_2]}{[1-(p_1-p_2)^2-(p_3-p_4)^2]^2-4(p_1-p_2)^2(p_3-p_4)^2}
\end{equation}
and
\begin{equation}
   \label{eq:Mzz}
   M_{zz}=1-4\Bigg(\frac{|u|^2}{a+d}+\frac{|v|^2}{b+c}\Bigg).
\end{equation}
It is seen from Eqs.~(\ref{eq:Mxx}) and (\ref{eq:Myy}) that always $M_{xx}\ge M_{yy}$
and therefore
\begin{equation}
   \label{eq:FF}
   {\cal F}=1-\max{\{M_{xx},M_{zz}\}}.
\end{equation}
Thus, this equation, together with expressions (\ref{eq:Mxx}) and (\ref{eq:Mzz}), is a
closed formula for the LQFI directly in terms of matrix elements and eigenvalues of an
arbitrary X-state.

Further, using Eqs.~(\ref{eq:avT}), (\ref{eq:avTm}) and relations $p_i=\exp(-E_i/T)/Z$,
we get expressions for the branches ${\cal F}_0=1-M_{zz}$ and ${\cal F}_1=1-M_{xx}$
via model parameters:
\begin{equation}
   \label{eq:F0}
   {\cal F}_0=\frac{2}{Z}\Big[\Big(\frac{r_1}{R_1}\Big)^{\!\!2}\sinh\frac{R_1}{T}\tanh\frac{R_1}{T}e^{-J_z/T}
	 +\Big(\frac{r_2}{R_2}\Big)^{\!\!2}\sinh\frac{R_2}{T}\tanh\frac{R_2}{T}e^{J_z/T}
	\Big]
\end{equation}
and
\begin{eqnarray}
   \label{eq:F1}
   &&{\cal F}_1=1-\frac{4}{Z}\Big(\cosh\frac{R_1}{T}e^{J_z/T}+\cosh\frac{R_2}{T}e^{-J_z/T}\Big)\Big(\cosh\frac{2J_z}{T}
	 \nonumber\\
	 &&+\cosh\frac{R_1}{T}\cosh\frac{R_2}{T}
	 +\frac{r_1r_2+B_2^2-B_1^2}{R_1R_2}\sinh\frac{R_1}{T}\sinh\frac{R_2}{T}\Big)/\Big[\Big(\cosh\frac{2J_z}{T}
	 \nonumber\\
	 &&+\cosh\frac{R_1}{T}\cosh\frac{R_2}{T}\Big)^2-\sinh^2\frac{R_1}{T}\sinh^2\frac{R_2}{T}\Big].
\end{eqnarray}
So, LQFI is given as
\begin{equation}
   \label{eq:F0F1}
   {\cal F}=\min{\{{\cal F}_0,{\cal F}_1\}}.
\end{equation}
Equations~(\ref{eq:F0})--(\ref{eq:F0F1}) together give an analytical expression for the
LQFI of the system (\ref{eq:H}) in a state of thermal equilibrium.

\section{
Local quantum uncertainty
}
\label{sect:LQU}
The LQU measure of quantum correlation, $\cal U$,
is based on the Wigner-Yanase skew information $\cal I$ \cite{WY63,L03}.
The LQU with respect to subsystem $A$, optimized over all local observables on $A$, is
defined as \cite{GTA13}
\begin{equation}
   \label{eq:Udef}
   {\cal U}(\varrho)=\min_{H_A}{\cal I}(\varrho,H_A).
\end{equation}
Thus, it is defined as the minimum quantum uncertainty associated to a single
measurement on one subsystem of bipartite system $AB$.
It is worth mentioning that the LQU is a genuine quantifier of quantum correlations,
and it has been shown that the LQU meets all the physical conditions of a criterion of
quantum correlations.

Importantly, the authors \cite{GTA13} were able to perform optimization for
qubit-qudit systems and presented the measure (\ref{eq:Udef}) in the form
\begin{equation}
   \label{eq:Um}
   {\cal U}=1-\lambda_{max}^{(W)},
\end{equation}
where $\lambda_{max}^{(W)}$ denotes the maximum eigenvalue of the $3\times3$
symmetric matrix $W$ whose entries are
\begin{equation}
   \label{eq:W}
   W_{\mu \nu}={\rm tr}\{\varrho^{1/2}(\sigma_\mu\otimes{\rm I})\varrho^{1/2}(\sigma_\nu\otimes{\rm I})\},
\end{equation}
with $\mu,\nu=x,y,z$ and $\sigma_{x,y,z}$ are the Pauli matrices as before.

Using Eqs.~(\ref{eq:RPvarrhoPR}) and (\ref{eq:RPxPR})--(\ref{eq:RPzPR}), we find that
the matrix $W$ for the system under consideration is diagonal and its eigenvalues are
given by
\begin{equation}
   \label{eq:Wxx}
   W_{xx}=2\frac{(\sqrt{p_1p_3}+\sqrt{p_2p_4})(q_1q_2+|uv|)^2+(\sqrt{p_1p_4}+\sqrt{p_2p_3})(q_1|v|-q_2|u|)^2}{(q_1^2+|u|^2)(q_2^2+|v|^2)},
\end{equation}
\begin{equation}
   \label{eq:Wyy}
   W_{yy}=2\frac{(\sqrt{p_1p_3}+\sqrt{p_2p_4})(q_1q_2-|uv|)^2+(\sqrt{p_1p_4}+\sqrt{p_2p_3})(q_1|v|+q_2|u|)^2}{(q_1^2+|u|^2)(q_2^2+|v|^2)}
\end{equation}
and
\begin{equation}
   \label{eq:Wzz}
   W_{zz}=\frac{q_1^2\Big[8|u|^2\sqrt{p_1p_2}+(a-d)^2(p_1+p_2)\Big]}{(q_1^2+|u|^2)^2}
	 +\frac{q_2^2\Big[8|v|^2\sqrt{p_3p_4}+(b-c)^2(p_3+p_4)\Big]}{(q_2^2+|v|^2)^2},
\end{equation}
where $q_1$ and $q_2$ are given by Eq.~(\ref{eq:q}) and $p_1,...,p_4$ are defined by
Eq.~(\ref{eq:p_i}).

Further calculations lead to \cite{JBD17,SDL18,HKD18,GPTTC21}:
\begin{equation}
   \label{eq:Wxx1}
   W_{xx}=(\sqrt{p_1}+\sqrt{p_2})(\sqrt{p_3}+\sqrt{p_4})+\frac{(b-c)(d-a)+4|uv|}{(\sqrt{p_1}+\sqrt{p_2})(\sqrt{p_3}+\sqrt{p_4})},
\end{equation}
\begin{equation}
   \label{eq:Wyy1}
   W_{yy}=(\sqrt{p_1}+\sqrt{p_2})(\sqrt{p_3}+\sqrt{p_4})+\frac{(b-c)(d-a)-4|uv|}{(\sqrt{p_1}+\sqrt{p_2})(\sqrt{p_3}+\sqrt{p_4})}
\end{equation}
and
\begin{equation}
   \label{eq:Wzz3}
   W_{zz}=\frac{1}{2}\Bigg[(\sqrt{p_1}+\sqrt{p_2})^2+(\sqrt{p_3}+\sqrt{p_4})^2
   +\frac{(d-a)^2-4|u|^2}{(\sqrt{p_1}+\sqrt{p_2})^2}
   +\frac{(b-c)^2-4|v|^2}{(\sqrt{p_3}+\sqrt{p_4})^2}\Bigg].
	 \nonumber\\
\end{equation}
It is clear from these expressions that $W_{xx}\ge W_{yy}$, so LQU formula for the
general two-qubit X states is
\begin{equation}
   \label{eq:UU}
   {\cal U}=1-\max{\{W_{xx},W_{zz}\}}.
\end{equation}

Using, as for LQFI, the expressions for the matrix elements and eigenvalues of the
density matrix $\varrho$, we arrive at formulas for the branches ${\cal U}_0=1-W_{zz}$
and ${\cal U}_1=1-W_{xx}$:
\begin{equation}
   \label{eq:U0}
   {\cal U}_0=\frac{4}{Z}\Big[\Big(\frac{r_1}{R_1}\Big)^{\!\!2}\sinh^2\frac{R_1}{2T}e^{-J_z/T}
	 +\Big(\frac{r_2}{R_2}\Big)^{\!\!2}\sinh^2\frac{R_2}{2T}e^{J_z/T}\Big]
\end{equation}
and
\begin{equation}
   \label{eq:U1}
   {\cal U}_1=1-\frac{4}{Z}\Bigg(\cosh\frac{R_1}{2T}\cosh\frac{R_2}{2T}
	 +\frac{r_1r_2+B_2^2-B_1^2}{R_1R_2}\sinh\frac{R_1}{2T}\sinh\frac{R_2}{2T}\Bigg).
\end{equation}
So,
\begin{equation}
   \label{eq:U0U1}
   {\cal U}=\min{\{{\cal U}_0,{\cal U}_1\}}.
\end{equation}
Equations~(\ref{eq:U0})--(\ref{eq:U0U1}) also give an explicit expression for the LQU
of the system (\ref{eq:H}) in a state of thermal equilibrium.

\section{
Discussion
}
\label{sect:Discus}
In the absence of external magnetic fields, $B_1=B_2=0$, the system (\ref{eq:H})
contains only two-particle interactions: both the Heisenberg XYZ and DM-KSEA
couplings.
Its Gibbs density matrix belongs to the general Bell-diagonal state.
Recently, the behavior of LQFI and LQU in this limiting case has been studied in detail
\cite{FY22}.
In particular, it was found that the branches of both LQFI and LQU are separated the
same boundary
\begin{equation}
   \label{eq:Bd-b}
   r_1+r_2=2|J_z|.
\end{equation}
Thus, in the case of $J_z\ne0$, the phase diagram in the parameter space is the plane
$(r_1,r_2)$; the quantity $|J_z|$ plays the role of a normalized constant and can be
considered as a unit without loss of generality.
Note that the boundary does not depend on temperature.
As a result, LQFI and LQU do not experience abrupt transitions with temperature
changes.

The picture becomes essentially different in the transition to general X-states.
For $B_1\ne B_2$, the quantum correlations, generally speaking, depend on which qubit
the measurement was performed;
the asymmetry in $B_1$ and $B_2$ is clearly visible for branches ${\cal F}_1$,
Eq.~(\ref{eq:F1}), and ${\cal U}_1$, Eq.~(\ref{eq:U1}).
Moreover, the parameter space is now five-dimensional, the boundaries, defined by
equations ${\cal F}_0={\cal F}_1$ and ${\cal U}_0={\cal U}_1$, are different and
involved the temperature.
The latter circumstance leads to the fact that with a smooth change temperature,
transitions from one branch of functions (\ref{eq:F0F1}) and (\ref{eq:U0U1}) to another
become possible, which will cause abrupt changes in the behavior of quantum
correlations.

Taking into account the arising new difficulties, it is useful, when passing to general
X-states, first of all to clarify the properties of LQFI and LQU at high and low
temperatures.

\subsection{
High-temperature behavior
}
\label{subsect:Too}
From expression (\ref{eq:F0}) it follows that
\begin{equation}
   \label{eq:F0hT}
	 {\cal F}_0(T)|_{T\to\infty}\approx
   \frac{r_1^2+r_2^2}{2T^2}+\frac{(r_2^2-r_1^2)J_z}{2T^3}
   -\frac{(5r_1^2+3r_2^2)R_1^2+(3 r_1^2+5r_2^2)R_2^2}{24 T^4}.
\end{equation}
Thus, the deviation from the Bell-diagonal case starts only from the fourth-order term
$1/T^4$.
The second branch of LQFI, Eq.~(\ref{eq:F1}), behaves as
\begin{equation}
   \label{eq:F1hT}
	 {\cal F}_1(T)|_{T\to\infty}\approx\frac{4B_1^2+4J_z^2+(r_1-r_2)^2}{4T^2}
	 +\frac{(R_2^2-R_1^2)J_z}{2T^3}.
\end{equation}
Likewise, for the LQU branches according to Eqs.~(\ref{eq:U0}) and (\ref{eq:U1}) we
have:
\begin{equation}
   \label{eq:U0hT}
	 {\cal U}_0(T)|_{T\to\infty}\approx\frac{r_1^2+r_2^2}{4T^2}
	 +\frac{(r_2^2-r_1^2)J_z}{4T^3}-\frac{(2r_1^2+3r_2^2)R_1^2+(3r_1^2+2r_2^2)R_2^2}{48T^4}
\end{equation}
and
\begin{equation}
   \label{eq:U1hT}
	 {\cal U}_1(T)|_{T\to\infty}\approx\frac{4B_1^2+4J_z^2+(r_1-r_2)^2}{8T^2}
	 +\frac{(R_2^2-R_1^2)J_z}{4T^3}.
\end{equation}
So, the quantum correlations decay at high temperatures according to the law $1/T^2$.

\subsection{
Zero-temperature limit
}
\label{subsect:T0}
When $T\to0$,
\begin{equation}
   \label{eq:F0lT}
   {\cal F}_0(T)|_{T\to0}
   \approx\frac{(r_1/R_1)^2\exp[(R_1-J_z)/T]+(r_2/R_2)^2\exp[(R_2+J_z)/T]}{\exp[(R_1-J_z)/T]+\exp[(R_2+J_z)/T]}.
\end{equation}
Therefore, in the limit of zero temperature
\begin{equation}
   \label{eq:F00}
   {\cal F}_0|_{T=0}=
	 \cases{
      (r_1/R_1)^2, &if $R_1>R_2+2J_z$\cr
      (r_2/R_2)^2, &if $R_1<R_2+2J_z$\cr
   }.
\end{equation}
The same is valid for the branch  ${\cal U}_0$:
\begin{equation}
   \label{eq:U00}
   {\cal U}_0|_{T=0}=
	 \cases{
      (r_1/R_1)^2, &if $R_1>R_2+2J_z$\cr
      (r_2/R_2)^2, &if $R_1<R_2+2J_z$\cr
   }.
\end{equation}
On the other hand, the second branch ${\cal U}_1(T)$ at zero temperature reaches the
maximum possible value equal to one:
${\cal U}_1|_{T=0}=1$, if $R_1-R_2\ne2J_z$.

Now we turn to the study of quantum correlations for arbitrary values of the model
parameters.

\subsection{
Temperature dependence
}
\label{subsect:vsT}
Generally,  $J_z\ne0$ and, consequently,  $|J_z|$ can be taken as a normalized
constant.
In this case, it is enough to consider functions
${\cal F}={\cal F}(T; r_1, r_2, B_1, B_2)$ and 
${\cal U}={\cal U}(T; r_1, r_2, B_1, B_2)$ for $J_z=1$ and $-1$.

We performed calculations on the both quantum correlations for various choices of model
parameters, trying to find and classify different types of their behavior.

\begin{figure}[t]
\begin{center}
\epsfig{file=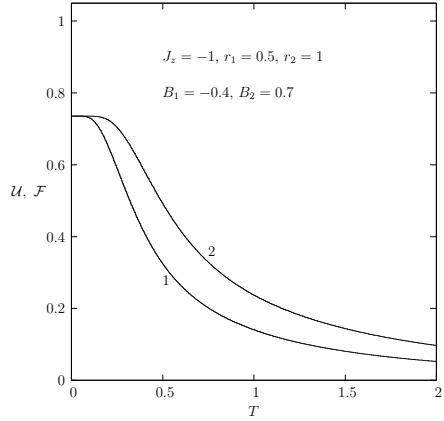,width=5.7cm}
\end{center}
\begin{center}
\caption{
Quantum correlations $\cal U$ (line 1) and $\cal F$ (line 2) versus temperature $T$.
The curves are given by 0-branches.
At $T=0$, both ${\cal U}$ and ${\cal F}$ are equal to $(r_1/R_1)^2=0.735294\ldots$
}
\label{fig:z2-5b}
\end{center}
\end{figure}

Figure~\ref{fig:z2-5b} shows the dependencies of ${\cal F}(T)$ and ${\cal U}(T)$ by
ferromagnetic coupling $J_z=-1$ and $r_1=0.5$, $r_2=1$, $B_1=-0.4$, and $B_2=0.7$.
%
%
Here, $R_1=0.583095$ and $R_2=1.486607$ and hence $R_1>R_2+2J_z$.
As can be seen from Fig.~\ref{fig:z2-5b}, the correlations decrease monotonically from
$(r_1/R_1)^2=0.735294$, which is determined in accordance with
Eqs.~(\ref{eq:F00}) and (\ref{eq:U00}),
to zero when the temperature increases from zero to infinity.
Both curves are defined by the branches ${\cal F}_0(T)$ and ${\cal U}_0(T)$, and no
sharp changes in the behavior of correlations are observed over the entire temperature
range, $T\in[0,\infty)$.

\begin{figure}[t]
\begin{center}
\epsfig{file=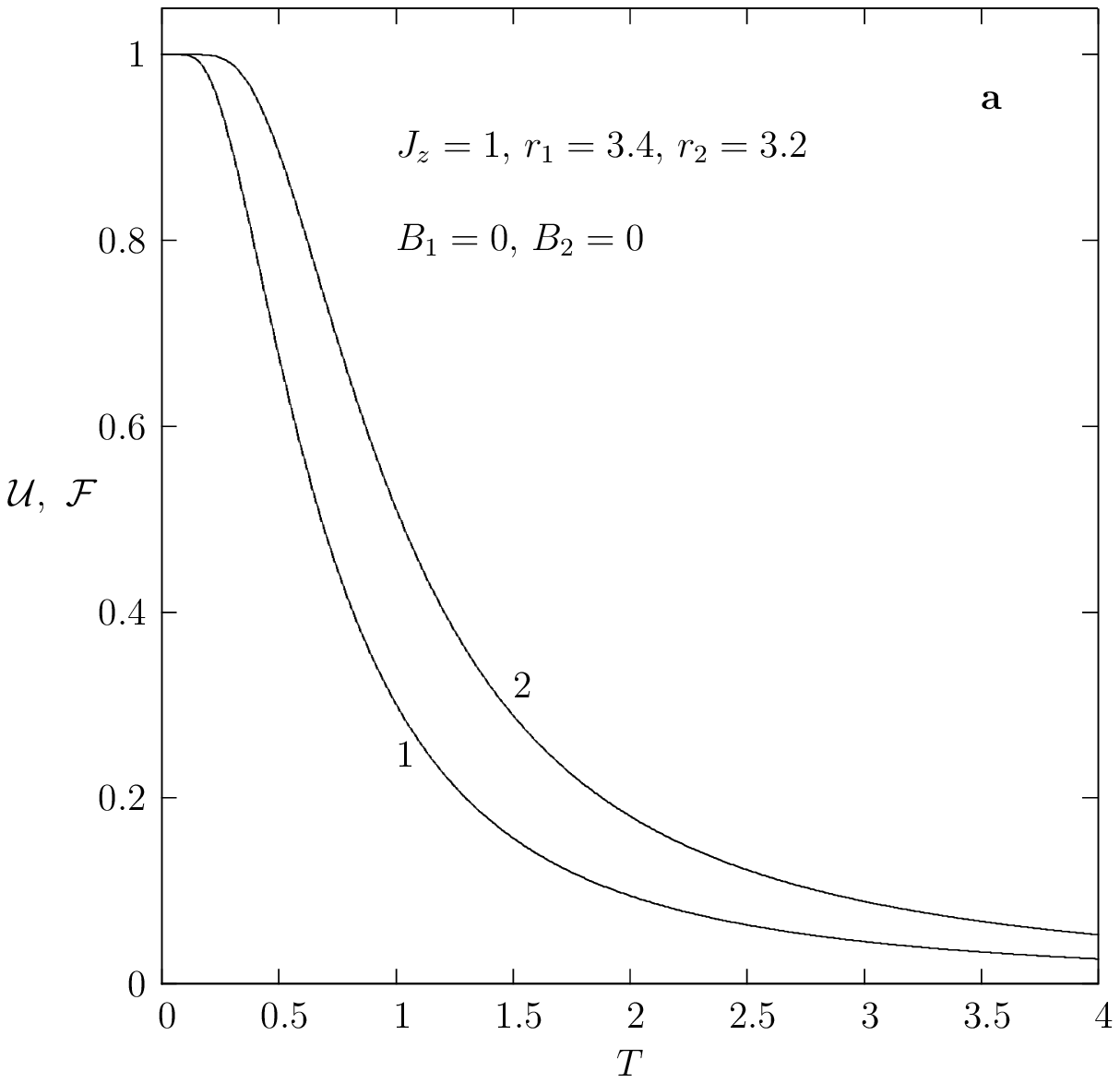,width=5.7cm}
\hspace{2mm}
\epsfig{file=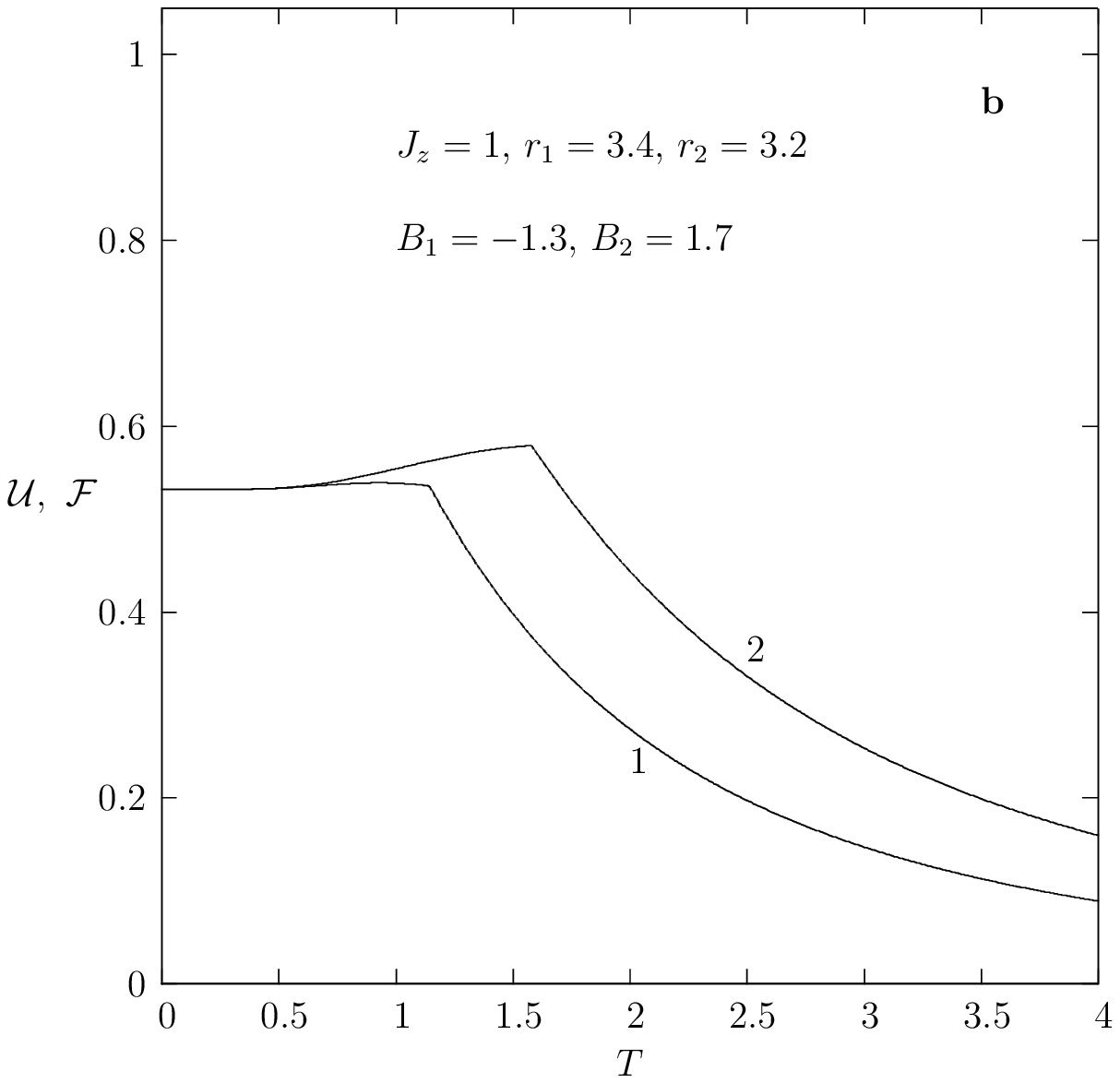,width=5.7cm}
\end{center}
\begin{center}
\caption{
Quantum correlations $\cal U$ (line 1) and $\cal F$ (line 2) versus the temperature
$T$.
($\rm\bf a$) -- External magnetic field is zero.
$\cal U$ and $\cal F$ are the branches ${\cal U}_1$ and ${\cal F}_1$, respectively.
The correlations are equal to one at $T=0$.
($\rm\bf b$) -- Nonzero external magnetic field.
The curves 1 and 2 have fractures at the temperatures respectively $1.1458$ and
$1.5821$,
where the quantum correlations suddenly pass from 1- to 0-branches.
At zero temperature, both correlations equal $(r_2/R_2)^2=0.5322245\ldots$
}
\label{fig:z2-1ab}
\end{center}
\end{figure}

Let us now put $J_z=1$ (anti-ferromagnetic coupling), $r_1=3.4$, $r_2=3.2$ and compare
the behavior of correlations in the absence of an external magnetic field and at
$B_1=-1.3$ and $B_2=1.7$.
Look at Fig.~\ref{fig:z2-1ab}.
%
%
As long as there is no external magnetic field, the functions ${\cal U}(T)$ and
${\cal F}(T)$ are smooth.
This is clearly seen in Fig.~\ref{fig:z2-1ab}$\rm\bf a$.
However, the picture changes dramatically when the system is exposed to magnetic
fields (see Fig.~\ref{fig:z2-1ab}$\rm\bf b$).
At high temperatures, both quantum correlations are branches  ${\cal U}_1(T)$ and
${\cal F}_1(T)$, as in the case of the absence of an external field.
But as the system cools, quantum correlations undergo sudden transitions in the
presence of external fields,
at first LQFI at the temperature $T=1.5821$ and then LQU at $T=1.1458$.
Correlation branches at these points change from  ${\cal F}_1$ and ${\cal U}_1$ to
${\cal F}_0$ and ${\cal U}_0$.
The curves remain continuous, but fractures are observed at the indicated points.
Finally, at absolute zero temperature, the values of both quantum correlations tend to
$(r_2/R_2)^2=0.5322245$, which is in full agreement with predictions (\ref{eq:F00}) and 
(\ref{eq:U00}).

\begin{figure}[t]
\begin{center}
\epsfig{file=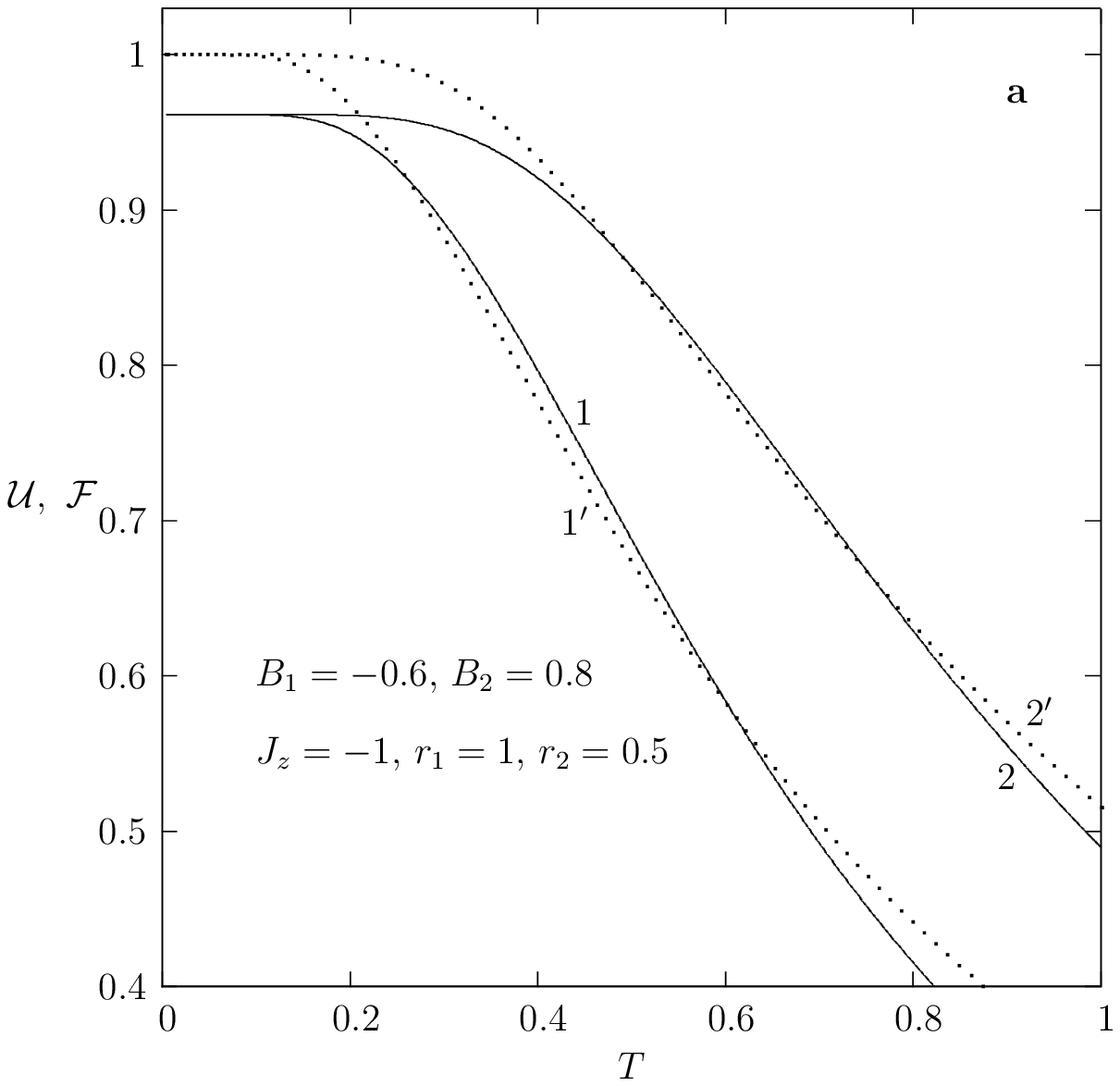,width=5.7cm}
\hspace{2mm}
\epsfig{file=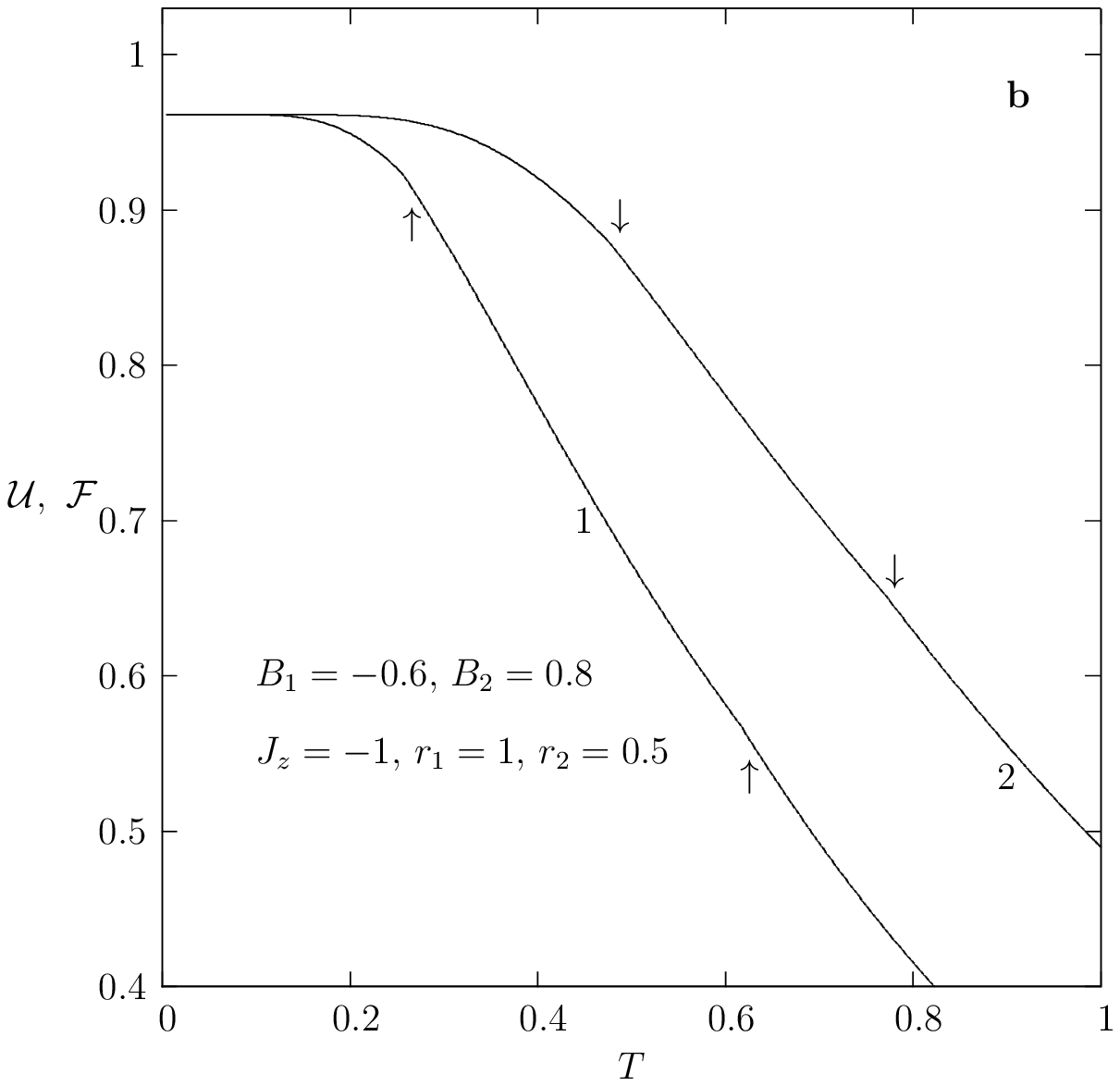,width=5.7cm}
\end{center}
\begin{center}
\caption{
Quantum correlations $\cal U$ and $\cal F$ and their branches depending on the
temperature $T$.
($\rm\bf a$) -- 
Lines 1 and 1$^\prime$ are branches ${\cal U}_0$ and ${\cal U}_1$, respectively;
they intersect at temperatures of 0.2565 and 0.6158.
Lines 2 and 2$^\prime$ are branches ${\cal F}_0$ and ${\cal F}_1$, respectively;
they intersect at temperatures 0.4778 and 0.7708.
($\rm\bf b$) --
correlations $\cal U$ (line 1) and $\cal F$ (line 2).
Arrows-up show the sudden change points for $\cal U$, and arrows-down mark similar
points for $\cal F$.
At $T=0$, both $\cal U$ and $\cal F$ are $(r_1/R_1)^2=0.961538\ldots$
}
\label{fig:z2-4ab}
\end{center}
\end{figure}

We discover another interesting phenomenon with a specific choice of model
parameters, namely, the presence of more than one abrupt transition.
Indeed, set, for definiteness, $J_z=-1$, $r_1=1$, $r_2=0.5$, $B_1=-0.6$ and $B_2=0.8$.
Figure~\ref{fig:z2-4ab}$\rm\bf a$ depicts the behavior of functions 
${\cal U}_0(T)$, ${\cal U}_1(T)$,  ${\cal F}_0(T)$, and ${\cal F}_1(T)$.
Each of the pairs of curves 1,~1$^\prime$ and 2,~2$^\prime$ intersects at two
points.
This leads to the fact that the quantum correlations
${\cal U}=\min\{{\cal U}_0,{\cal U}_1\}$ and ${\cal F}=\min\{{\cal F}_0,{\cal F}_1\}$
experience {\em two} sudden transitions each (see Fig.~\ref{fig:z2-4ab}$\rm\bf b$).
%
%

\subsection{
Dependence on an external field
}
\label{subsect:vsB}
It was shown in Ref.~\cite{FY22} that LQFI and LQU in the system (\ref{eq:H}) without a
magnetic field can undergo abrupt transitions as the exchange constants smoothly vary.
Therefore, it is not surprising that abrupt transitions occur in the entire system
(\ref{eq:H}) when the fields $B_1$ and $B_2$ change.
We illustrate this on an example.

\begin{figure}[t]
\begin{center}
\epsfig{file=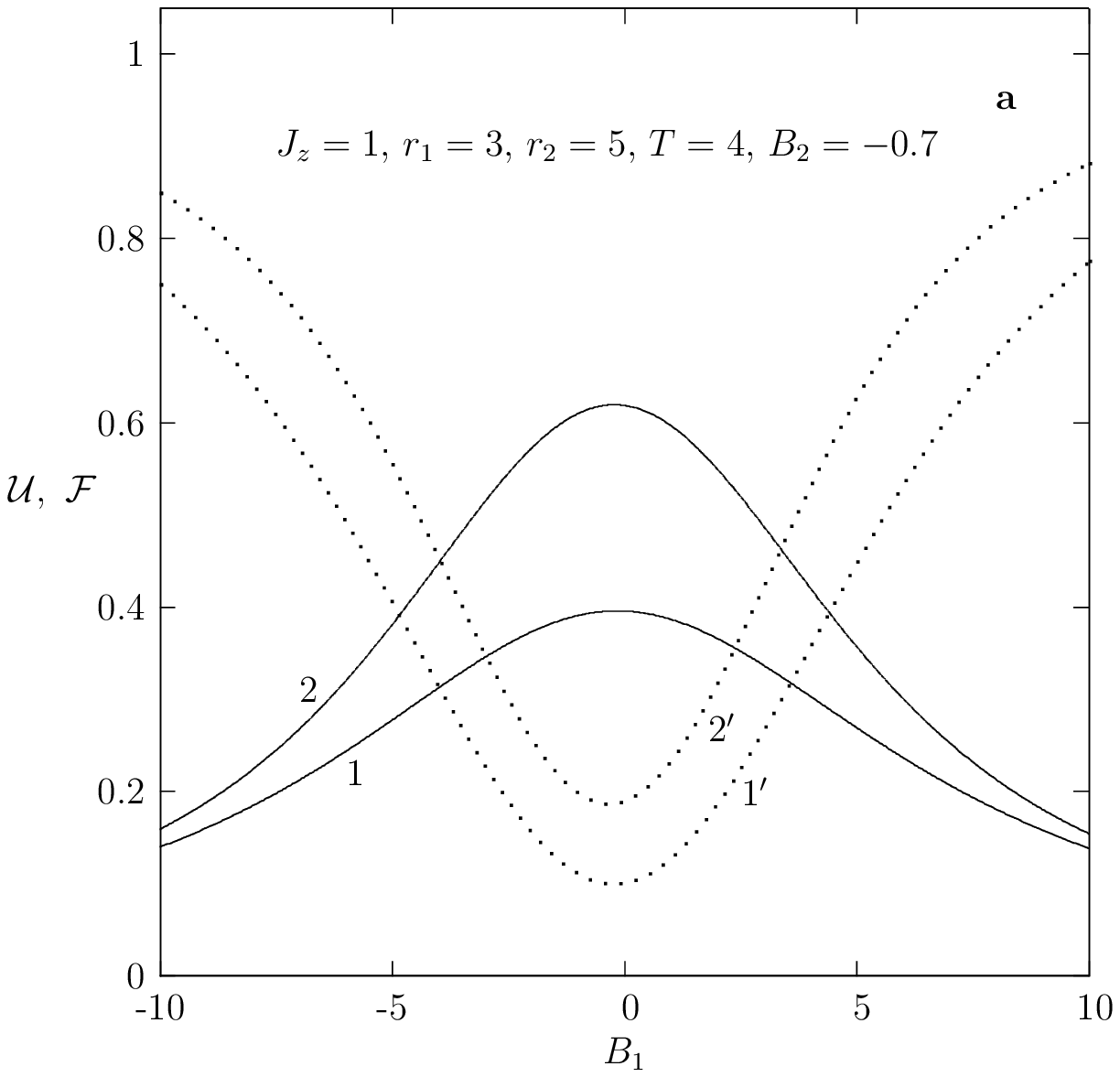,width=5.7cm}
\hspace{2mm}
\epsfig{file=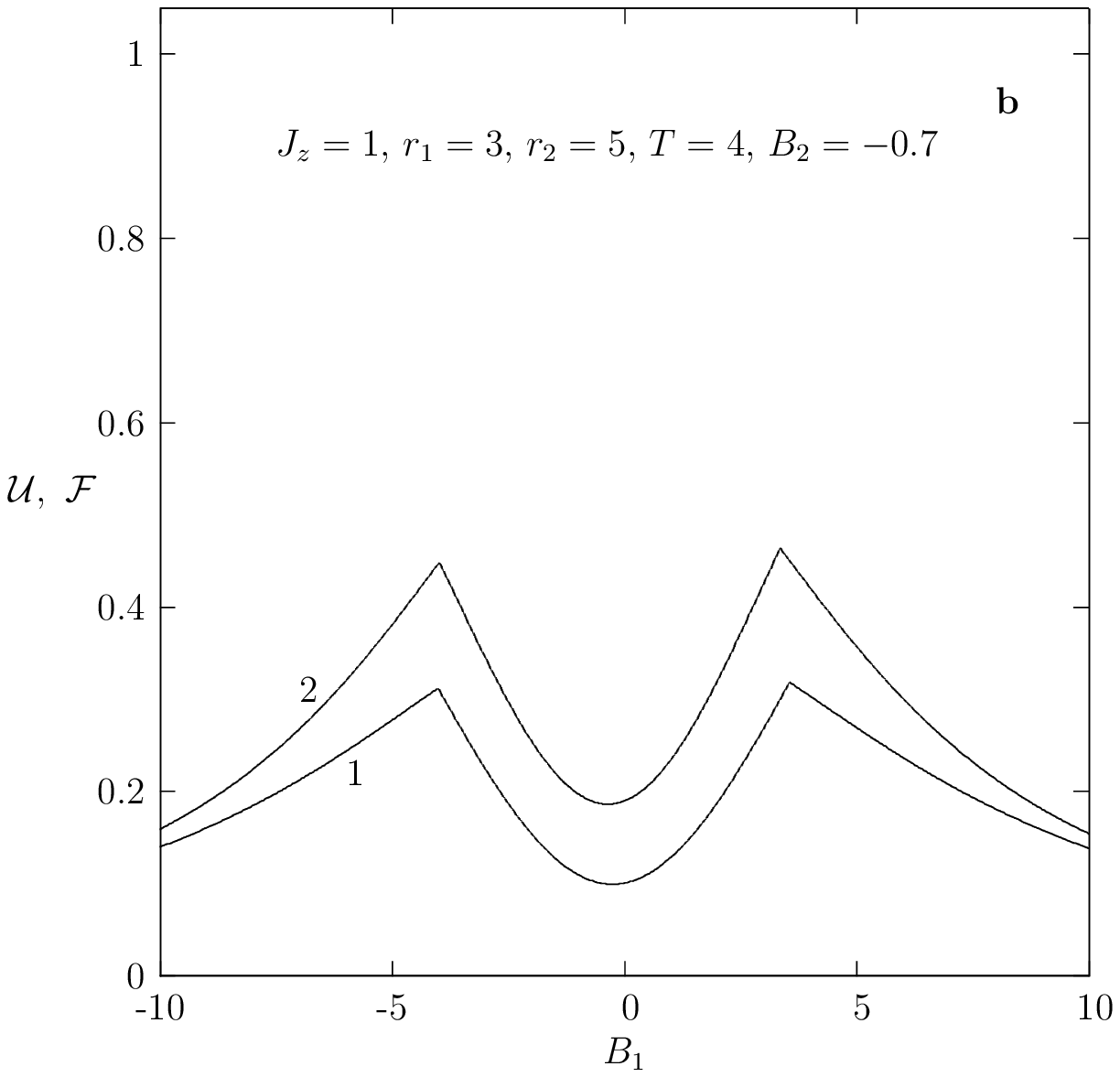,width=5.7cm}
\end{center}
\begin{center}
\caption{
Quantum correlations ${\cal U}$ and ${\cal F}$ and their branches versus the external
magnetic field $B_1$.
($\rm\bf a$) --
Solid lines 1 and 2 are the branches ${\cal U}_0$ and ${\cal F}_0$, and
dotted lines 1$^\prime$ and 2$^\prime$ are the branches ${\cal U}_1$ and ${\cal F}_1$,
respectively.
($\rm\bf b$) --
Correlations $\cal U$ (lines 1) and $\cal F$ (lines 2)
}
\label{fig:z3-1ab}
\end{center}
\end{figure}

Take the following set of fixed parameters: $J_z=1$, $r_1=3$, $r_2=5$, $B_2=-0.7$ and
$T=4$, and consider the behavior of quantum correlations and their individual branches
as a function of an external magnetic field $B_1$.
Figure~\ref{fig:z3-1ab}$\rm\bf a$ discloses the mechanism of appearance of fractures
on the curves of quantum correlations.
%
%
The reason again lies in the intersection of different branches.
Together with the minimization conditions, required by Eqs.~(\ref{eq:F0F1}) and
(\ref{eq:U0U1}), this leads to piecewise-defined functions which branches are separated 
by points of sharp changes.
The fractures on the curves of both quantum correlations are clearly seen
in Fig.~\ref{fig:z3-1ab}$\rm\bf b$.

\section{
Summary and outlook
}
\label{sect:Concl}
In this paper, the two-qubit Heisenberg XYZ system influenced by both antisymmetric
Dzyaloshinsky--Moriya and symmetric Kaplan--Shekhtman--Entin-Wohlman--Aharony
interactions and exposed to an external magnetic field has been considered at thermal
equilibrium.
For that, we have examined the behavior of two measures of discord-like quantum
correlation, namely, the LQFI and LQU.

Both measures have been shown to exhibit sudden transitions with a smooth change in
temperature, which are absent in the same spin systems, but without an external field.
In addition, it has been found that quantum correlations can sequentially experience
several such transitions.

Points of abrupt changes in the behavior of LQFI and LQU do not coincide for the
general X state.
As a result, in contrast to the Bell-diagonal quantum states, the quantitative
agreement is now poorer, although the qualitative behavior of both correlations is
still preserved.

The derived formulas for LQFI and LQU can also be useful in other problems with general
two-qubit X-states,
for example, when studying the dynamics of decoherence in various quantum systems.



\end{document}